# A Novel Near-field Photonic Thermal Diode with hBN and InSb

Dudong Feng, Shannon K. Yee, and Zhuomin M. Zhang*

George W. Woodruff School of Mechanical Engineering,
Georgia Institute of Technology, Atlanta, GA 30332, USA

*Corresponding Author: zhuomin.zhang@me.gatech.edu

**ABSTRACT:** Similar to the diode in electronics, a thermal diode is a two-terminal device that allows heat to transfer easier in one direction (forward bias) than in the opposite direction (reverse bias). Unlike conductive and convective thermal diodes, a photonic thermal diode operates in a contactless mode and may afford a large operating temperature range. In this work, a novel near-field photonic thermal diode with hexagonal boron nitride (hBN) and indium antimonide (InSb) is proposed and theoretically demonstrated. The temperature dependence of the interband absorption of InSb is used to couple (or decouple) with the hyperbolic phonon polaritons in hBN. The numerical analysis predicts a rectification ratio greater than 17 for a 10 nm vacuum gap when operating at an average temperature of 300 K and a temperature difference of 200 K. The calculated rectification ratio exceeds 35 with higher average temperatures and larger temperature differences. The mechanism proposed here for achieving photonic thermal rectification provides a new way of controlling radiative heat transfer.

Analogous to its electronic counterpart, a thermal diode is a two-terminal device with an asymmetric thermal transfer function that depends on the direction of temperature gradient.[1-3] Photonic thermal diodes based on radiative heat transfer between the two terminals at different temperatures separated by a vacuum gap are enabled by the temperature-dependent dielectric functions (or emissivities in the far field) of the materials used.[4-8] Photonic thermal diodes have advantages over phononic thermal diodes in terms of operating in a noncontact mode and over a relatively large temperature range. The rectification ratio is often used to quantify the performance of a thermal diode, and it is defined as $R = q_\mathrm{f}/q_\mathrm{r} - 1$, where $q_\mathrm{f}$ and $q_\mathrm{r}$ represent the magnitude of the forward and reverse heat fluxes, respectively. Researchers are seeking thermal diode designs with high rectification ratios for possible applications in the energy conversion, electronic thermal management, building thermal regulation, and human thermal comfort.[9,10]

Radiative heat transfer can be greatly enhanced in the near field (when the separation distance between the two objects are smaller than the thermal radiation wavelengths), especially with micro/nanostructures or metamaterials, and has been extensively investigated in recent years both theoretically and experimentally.[10-15] Near-field thermal modulation and rectification have also been demonstrated experimentally by several groups.[16-19] Otey et al. proposed the first photonic thermal rectification scheme based on the temperature dependence of electromagnetic resonances in different SiC polytypes and predicted a rectification ratio of 0.41 at a 100 nm vacuum gap with a temperature difference of 200 K.[4] Wang and Zhang theoretically investigated thermal rectification between dissimilar materials and predicted $R = 9.9$ with a 5 nm vacuum gap between Si and $SiO_2$ at temperatures of 1000 K and 300 K, respectively.[6] Thermochromic materials and metal-to-insulator phase change materials (PCMs) such as $VO_2$ have been extensively studied for use in photonic thermal diode in both the near field[20,21] and the far field.[22-25] Graphene-coated structures have also been considered to build photonic thermal diodes.[26,27] The best estimated rectification ratios with planar or multilayer films are around 10 as summarized by Wen et al.[28] Furthermore, asymmetrically patterned photonic nanostructures have been considered to achieve higher rectification ratio ($R > 10$).[29] By coupling a PCM (VO$_2$) with gratings and multilayered nanostructures, Chen et al. numerically demonstrated $R \approx 24$ at 50 nm vacuum gap with a temperature difference of 10 K above and below the VO$_2$ transition temperature of 341 K.[30] High rectification ratios may also be achieved with superconductors but the operating temperatures are limited to below 20 K.[31,32]

In this work, a new scheme of near-field photonic thermal diode is proposed with a semiconductor InSb film and a hexagonal boron nitride (hBN) sheet. The underlying mechanisms is to couple/decouple the interband absorption of InSb with the hyperbolic phonon polariton (HPhP) of hBN in the mid-infrared region. InSb is a narrowband semiconductor material with a bandgap energy $E_g \approx 0.175 \text{ eV}$ (corresponding to a wavelength of $7.1 \text{ μm}$) and has been considered for near-field thermophotovoltaic applications.[33,34] The bandgap energy increases as the temperature is lowered, resulting in a shift of the interband absorption edge toward short wavelengths.[35,36] In recent years, hyperbolic metamaterials have been extensively studied for subwavelength imaging and near-field radiative transfer.[12-14,37-39] Some naturally occurring 2D materials with layered structures (such as hBN) exhibit hyperbolic bands and support HPhPs.[40] Near-field radiative heat transfer can greatly exceed the blackbody limit at nanoscale separation distances when HPhPs are excited in the hBN sheets.[41,42] The strong temperature dependence of the absorption edge allows the absorption of InSb to couple (or decouple) with the hyperbolic band of hBN, when the temperature of InSb is higher (or lower) than that of hBN. The new scheme proposed here may enable higher rectification ratio without using gratings or nanoparticles. Furthermore, the photonic thermal diode based on InSb and hBN can operate in a relatively large temperature range.

**RESULTS AND DISCUSSION**

As shown in Figure 1a, the photonic thermal diode consists of an InSb film whose thickness is $h_1$ at temperature $T_1$ and a hBN sheet whose thickness is $h_2$ at temperature $T_2$, separated by a vacuum gap of distance $d$. The medium behind InSb (or hBN) is assumed to be vacuum that extends to infinity and is at a thermal equilibrium with the adjacent solid. The net heat flux from media 1 to 2 is represented by $q_{12}$. For the structure considered here, the forward bias is for $T_1 > T_2$ so that $q_f = q_{12}$. For reverse bias ($T_1 < T_2$), $q_r = -q_{12}$ because $q_{12}$ is negative. The dielectric functions of both hBN and InSb are needed for computing the radiative heat flux.

The hBN sheet is a uniaxial medium due to its layered structures. In the present study, the optical axis of hBN is parallel to $z$-direction so that its dielectric tensor may be expressed as

$$\bar{\bar{\varepsilon}}_2 = \begin{pmatrix} \varepsilon_\perp & 0 & 0 \\ 0 & \varepsilon_\perp & 0 \\ 0 & 0 & \varepsilon_\parallel \end{pmatrix} \tag{1}$$

where $\varepsilon_\perp$ is the ordinary dielectric function (when the electric field is in the $x$-$y$ plane) and $\varepsilon_\parallel$ is the extraordinary dielectric function (when the electric field is parallel to the optical axis or $z$-axis). Lattice vibrations in either the in-plane or out-of-plane direction result in a mid-infrared Reststrahlen band in the ordinary or extraordinary dielectric function, respectively. Therefore, the dielectric functions of hBN can be modeled as[40,41]

$$\varepsilon_m(\omega) = \varepsilon_{\infty,m}\left(1 + \frac{\omega_{LO,m}^2 - \omega_{TO,m}^2}{\omega_{TO,m}^2 - i\gamma_m\omega - \omega^2}\right) \tag{2}$$

where $\omega$ is the angular frequency, $m = \perp$ or $\parallel$, $\varepsilon_\infty$ is a high-frequency constant, TO and LO represent transverse optical phonon and longitudinal optical phonon modes, respectively, and $\gamma$ is the damping coefficient. These parameters are taken from Kumar et al.[40] and assumed to be independent of temperature. The real parts of dielectric functions of hBN are shown in Figure 1b. There are two hyperbolic bands in which $\varepsilon'_\perp \varepsilon'_\parallel < 0$. The type II hyperbolic band with $\varepsilon'_\perp < 0$ in the region $0.17 \text{ eV} < \hbar\omega < 0.20 \text{ eV}$ is shaded because it plays a key role in the proposed thermal diode. As shown in Figure 1c, a large spike occurs in the imaginary part of the ordinary dielectric function ($\varepsilon''_\perp$) near 0.17 eV, and $\varepsilon''_\perp$ decays quickly toward higher and lower photon energies; this suggests a strong resonance at $\omega = \omega_{\text{TO},\perp}$. The real part of the extraordinary dielectric function ($\varepsilon'_\parallel$) is nearly a constant ($\approx 2.8$) while the imaginary part ($\varepsilon''_\parallel$) is negiligible small for $\hbar\omega > 0.15 \text{ eV}$.

As a widely used narrow bandgap semiconductor, the interband absorption coefficient of InSb is strongly temperature dependent.[35,36] For intrinsic semiconductors, the absorption due to free carriers and phonon vibartions is negiligible compared to the interband absorption from low to intermediatly temperatures. Therefore, the dielectric functions of updoped InSb can be approximatly modeled as[33]

$$\varepsilon_1(\omega,T) = \left( n_1 + i\frac{\alpha}{2k_0} \right)^2 \quad (3)$$

where $n_1$ is the refractive index, $k_0 = \omega/c$ is the wavevector with $c$ being the speed of light in vacuum, and $\alpha$ is the absorption coefficient due to interband absorption given by

$$\alpha(\omega,T) = \begin{cases} 0, & \omega < \omega_g(T) \\ \alpha_0 \left[ \dfrac{\omega - \omega_g(T)}{\omega_g(T)} \right]^{1/2}, & \omega > \omega_g(T) \end{cases} \quad (4)$$

and $\omega_g(T) = E_g(T)/\hbar$ is the angular frequency corresponding to the bandgap energy $E_g$ with $\hbar$ being the reduced Planck constant. It is reasonable to use the fitting value of $\alpha_0 \approx 7000 \text{ cm}^{-1}$ for all temperatures.[33,35] The temperature-dependent bandgap energy of InSb has been fitted to the Varshni relation as follows:

$$E_g(T) = 0.235 - 2.7 \times 10^{-4} T^2 / (T+106) \qquad (5)$$

where $T$ is in [K] and $E_g$ in [eV]. This relation has been experimentally validated from about 10 K to 600 K.[36,43] As shown in Figure 1d, the bandgap energy (the rising point of the curve) shifts to small value as the temperature increases. For InSb, the imaginary part of the dielectric function is proportional to the absorption coefficient if the refractive index is taken as a constant ($n_1 = 3.96$).[44] As the temperature increases, the interband absorption spectrum of hBN approaches and then overlaps with the hyperbolic band of hBN (shaded region in Figure 1d). Coupled with the HPhPs of hBN enables, photon tunneling can occur through the nanoscale vacuum gap in a narrow band when the InSb film is at a relatively higher temperature. The method to calculate the near-field radiative heat transfer can be found Liu et al.[12] and Zhang.[14] A brief discussion is provided under the METHOD section.

For $h_1 = 100$ μm (InSb), $h_2 = 5$ nm (hBN), and a vacuum gap of $d = 10$ nm, the heat flux $q_{12}$ is calculated with an average temperature $T_{avg} = (T_1 + T_2)/2$ as a function of $\Delta T = T_1 - T_2$ and plotted in Figure 2 for both forward and reverse bias case. Nanometer thickness of hBN is chosen to obtain strong resonance with sharp peaks in the heat flux spectrum at nanometer distances.[42] The curve resembles the current-voltage curve of a diode (dark current curve) with an exponential growth of the heat flux in the forward direction and a much smaller magnitude in the reverse direction. In contrary, for thermal diode made of VO₂, the heat flux changes linearly with temperature difference (though the slope is different between the forward and reverse biases).[30] In the reverse direction, the heat flux does not change significantly with the proposed design.

Therefore, increasing the temperature difference results in an increase of the rectification ratio. The rectification ratio is also shown in Figure 2 as a function of $|\Delta T|$ using the double $y$-axis plot. When $|\Delta T| = 200$ K, $R$ = 17.1, which surpasses previously reported figures in the similar temperature range for planar or multilayered structures at $d \geq 10$ nm. The heat flux with $T_1 = 400$ K and $T_2 = 200$ K is about 9.7 times that between two blackbodies at these temperatures. The near-field heat flux enhancement between InSb and hBN is not as strong as that between two hBN sheets.[41,42] When the heat flow direction is reversed ($T_1 = 200$ K; $T_2 = 400$ K), the heat flux is about 53% that between two blackbodies. The reduction is presumably due to multiple reflections with the InSb and hBN layers. Even with a smaller $|\Delta T|$, relatively large $R$ can still be achieved: for example, $R$ = 2.17, 5.13 and 9.67, when $|\Delta T|$ = 60 K, 92 K and 120 K, respectively. Note that another way to characterize the thermal performance is to use a rectification efficiency defined as $\eta = 1 - q_r / q_f$. A rectification ratio of 9.67 or 17.1 correspond to $\eta = 0.906$ and 0.945, respectively.

The spectral heat flux $q(\omega)$ for both forward and reverse scenarios are shown in Figure 3a when $\Delta T = 200$ K and the other conditions the same as for Figure 2. The spectral heat flux between two blackbodies at 400 K and 200 K are also shown for comparison. The forward $q(\omega)$ is significantly higher than that for the reverse case heat flux near the hyperbolic region. Since $E_g(400 \text{ K}) \approx 0.15$ eV, $q(\omega)$ for forward bias starts to increase at $\hbar\omega > 0.15$ eV due to frustrated mode modes,[14] which increases sharply at 0.17 eV where $\varepsilon''_\perp$ reaches a peak. As $\hbar\omega$ increases beyond 0.17 eV, it falls in the hyperbolic band of hBN where $q(\omega)$ continues to increase and reach a peak near 0.182 eV and then decreases as $\hbar\omega$ further increases. Interference within the InSb film causes oscillations in the heat flux spectra for $\hbar\omega < E_\text{g}$. When $\hbar\omega < 0.15$ eV, the dielectric function of InSb is the same at 200 K and 400 K, and the forward and backward spectra overlap with each other. Because InSb is treated as nonabsorbing in this region, only propagating waves in vacuum can be supported. Furthermore, surface reflection results in a reduction of the heat transfer that is even smaller than that between two blackbodies. The oscillations continue for reverse bias $E_\text{g}(200\text{ K}) \approx 0.2$ eV. A small peak occurs at 0.17 eV in the reverse scenario due to the spike in $\varepsilon_\perp''$. For reverse bias, there is a quick rise in $q(\omega)$ near 0.2 eV; however, the magnitude is relatively small since it is beyond the hBN hyperbolic band and only propagating waves in vacuum and frustrated modes can be supported. For $\hbar\omega > 0.23$ eV, the difference between the forward and reverse scenarios is negligibly small. In this region, the absorption by the thin hBN sheet is negligible so that only propagating waves in vacuum are supported. No interference effects are expected at $\hbar\omega > 0.20$ eV since InSb film is essentially opaque in this region. In this case, the surface reflection between vacuum and InSb is dominated by the refractive index rather than the absorption coefficient. Hence, the forward and backward $q(\omega)$ almost overlap with each other.

The contour plot of the transmission coefficient is shown in Figure 3b for forward bias to gain a better insight of the hyperbolic modes. The hyperbolic dispersion curve is observed in the region from 0.17 eV to 0.19 eV (though it extends to about 0.2 eV) where HPhPs are excited. At small $\beta$ values, there is a bright spot near 0.17 eV which is attributed to frustrated modes in InSb. As mentioned previously, these modes give photon tunneling for $\hbar\omega$ from 0.15 eV to 0.17 eV, although with a lower transmission coefficient away from 0.17 eV. There exist propagating waves in InSb that can be tunneled through the vacuum gap at $\beta < n_1 k_0$ due to frustrated total internal reflection.[14] The hyperbolic modes support photon tunneling at high $\beta$ values, resulting in greater enhancement of the near-field heat flux.[12] Due to the shift of the bandgap, HPhPs cannot be excited in the reverse biased scenario.

Parametric sweeps were conducted to help understand the role of thicknesses and vacuum gap on the performance of the proposed photonic thermal diode. The calculation results show that the effect of the InSb thickness is small. When $h_1 > 10$ μm, the heat fluxes and rectification ratio are almost constant. Therefore, $h_1 = 100$ μm is used in all calculations without further discussion. The effects of $d$ and $h_2$ on the rectification ratio are shown in Figure 4 for the two terminal temperatures of 400 K and 200 K. In general, reducing $d$ gives rise to the forward heat flux, resulting in a higher rectification ratio. From practical consideration, the smallest $d$ value is taken to be 10 nm. When $d$ is increased to 50 nm, the rectification ratio reduces to less than 2. As $h_2$ increases, $R$ increases and reaches a plateau for small $d$ values. For $d = 10$ nm, $R$ reaches a maximum at $h_2 = 5$ nm and decreases slightly as $h_2$ increases. Even when $h_2 = 1000$ nm, the rectification ratio is greater than 15.4. For $d = 20$ nm, the $R = 4.54$ (or rectification efficiency $\eta = 0.82$) at $h_2 = 10$ nm.

It can be seen from Figure 2 that the rectification ratio increases with the temperature difference. Furthermore, the average temperature may also affect the photonic thermal diode performance. In Figure 5, the forward and reverse heat fluxes and the rectification ratio are plotted as a function of the absolute temperature difference for $T_{avg} = 300$ K, 350 K, and 400 K. The upper and lower temperature limits are set as 50 K and 600 K, respectively. Therefore, the upper limit of $|\Delta T|$ is set to 500 K for $T_{avg} = 300$ K and 350 K and 400 K for $T_{avg} = 400$ K. As shown in Figure 1c, as the temperature increases, the band edge of InSb moves toward lower frequencies and starts to overlap with the hBN hyperbolic band. The coupling becomes stronger when the temperature of InSb increases to beyond 300 K where $E_g \approx 0.175$ eV. For $T_{avg} = 300$ K, the rectification ratio exceeds 10 when the lower temperature terminal is below about 240 K at which $E_g \approx 0.19$ eV is near the peak of the spectral heat flux as shown in Figure 3a. This explains why the rectification ratio is smaller for higher $T_{avg}$ when $|\Delta T|$ is relatively small. Crossovers occur as $|\Delta T|$ increases and the highest rectification ratio for $|\Delta T| = 400$ K occurs at $T_{avg} = 400$ K where $R = 36.7$ and $q_f = 1.57 \times 10^5$ W/m$^2$, which is 21.6 times that between two blackbodies at 600 K and 200 K. For $T_{avg} = 350$ K and $|\Delta T| = 500$ K, the rectification ratio is close to 38. When $T_1$ is increased from 400 K and 600 K, the absorption coefficient of InSb increases, which subsequently enhance the coupling with the hBN hyperbolic band. The rectification ratios obtained from this study exceed the literature values between planar as well as grating structures in the similar temperature range. Compared to the photonic thermal diode employing PCMs, the proposed device can afford a wider range of working temperatures with higher rectification ratios when operated at nanoscale vacuum gaps.

**CONCLUSION**

In summary, a novel near-field photonic thermal diode is proposed by coupling the interband absorption of an InSb film with the HPhPs in a hBN sheet. The temperature dependence of the band edge absorption of InSb enables the thermal diode behavior. A high rectification ratio of more than 17 is predicted at a 200 K temperature difference with an average temperature of 300 K. When the temperature difference is increased to 350 K for a temperature difference of 500 K, calculations show that the rectification ratio is near 38. This work theoretically demonstrates a new mechanism to achieve photonic thermal rectification and provides a new way of controlling the radiative heat transfer at the nanometer scales.

**METHODS**

The fluctuational electrodynamics and a transfer matrix method are used to calculate the near-field radiative heat transfer between two thin films. The net radiative heat flux from 1 to 2 is calculated by[12-14]

$$q_{12} = \frac{1}{4\pi^2} \int_0^\infty \left[\Theta(\omega,T_1) - \Theta(\omega,T_2)\right] d\omega \int_0^\infty \sum_{j=s,p} \xi_j(\omega,\beta) \beta d\beta \qquad (6)$$

where $\Theta(\omega,T) = \hbar\omega \left(e^{\hbar\omega/k_B T} - 1\right)^{-1}$ is the mean energy of the Planck oscillator, $T_1$ and $T_2$ denote respectively the temperatures of InSb and hBN as shown in Figure 1a, $\beta = \left(k_x^2 + k_y^2\right)^{1/2}$ is the magnitude of the wavevector in the *x-y* plane, and $\xi_j(\omega,\beta)$ is the energy transmission coefficient for either transverse electric (TE) waves (*s*-polarization) or transverse magnetic (TM) waves (*p*-polarization). The energy transmission coefficient is calculated from[14]

$$\xi_j(\omega,\beta) = \begin{cases} \dfrac{\left(1 - r_{1j} r_{1j}^*\right)\left(1 - r_{2j} r_{2j}^*\right)}{\left|1 - r_{1j} r_{2j} e^{2ik_{z0}d}\right|^2}, & \beta < k_0 \\[2mm] \dfrac{4\,\mathrm{Im}(r_{1j})\,\mathrm{Im}(r_{2j}) e^{-2|k_{z0}|d}}{\left|1 - r_{1j} r_{2j} e^{2ik_{z0}d}\right|^2}, & \beta > k_0 \end{cases} \qquad (7)$$

where $k_{z0} = \left(k_0^2 - \beta^2\right)^{1/2}$ is the *z*-component wavevector in vacuum, and $r_{1j}$ and $r_{2j}$ are the reflection coefficients of the semi-infinite media including InSb or hBN, respectively. They are calculated by treating hBN as a uniaxial thin film and InSb as an isotropic medium.[12]

**AUTHOR INFORMATION**

**Corresponding Author**

*E-mail: zhuomin.zhang@me.gatech.edu.

**Notes**

The authors declare no competing financial interest.

**ACKNOWLEDGMENTS**

This work was mainly supported by the U.S. Department of Energy, Office of Science, Basic Energy Sciences (DE-SC0018369). SKY would like to thank partial support from the Department of Navy award (N00014-19-1-2162) issued by the Office of Naval Research. ZMZ would also like to thank the support of the National Science Foundation (CBET-2029892).

**Figure captions**

Fig. 1.  (a) Schematic of the near-field radiative thermal diode with an InSb film (thickness $h_1$) and hBN sheet (thickness $h_2$) at temperatures of $T_1$ and $T_2$, respectively, separated by a vacuum gap of distance $d$. (b) Ordinary and extraordinary dielectric functions (real part only) of hBN, showing the two hyperbolic bands. (c) Dielectric functions of hBN near the type II hyperbolic band, noting that the imaginary part of the extraordinary component is negligibly small in this region. (d) The imaginary part of the dielectric function of InSb at different temperatures.

Fig. 2.  The net heat flux and rectification ratio versus the temperature difference for the proposed near-field photonic thermal diode calculated at an average temperature of 300 K for $h_1$ = 100 μm, $h_2$ = 5 nm, and $d$ = 10 nm.

Fig. 3.  (a) Spectral heat flux for both forward and reverse scenarios and between two blackbodies for $\Delta T$ = 200 K and $T_{avg}$ = 300 K; the other conditions are the same as for Figure 2. (b) Contour plot of the transmission coefficient under forward bias.

Fig. 4.  Rectification ratio versus the thickness of hBN film at different vacuum gap distances, when the thickness of InSb is fixed at 100 μm for $\Delta T$ = 200 K and $T_{avg}$ = 300 K.

Fig. 5.  (a) The net heat flux and (b) rectification ratios verse the temperature difference of the near-field photonic thermal diode at $T_{avg}$ of 300 K, 350 K, and 400 K; the other conditions are the same as in Figure 2.

**TOC Figure**

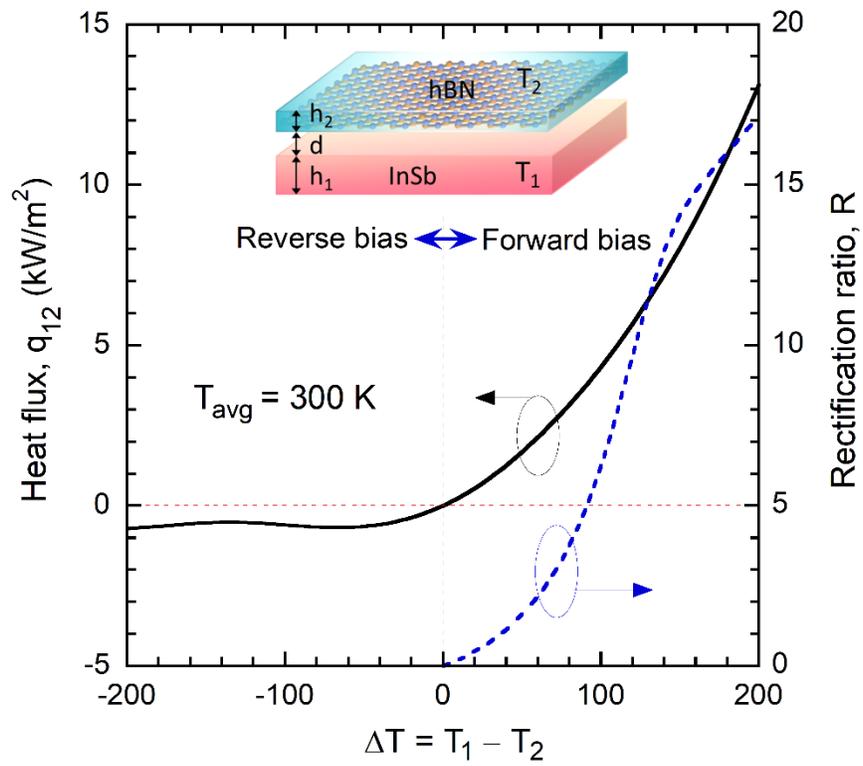

**TOC Figure**, Feng et al.

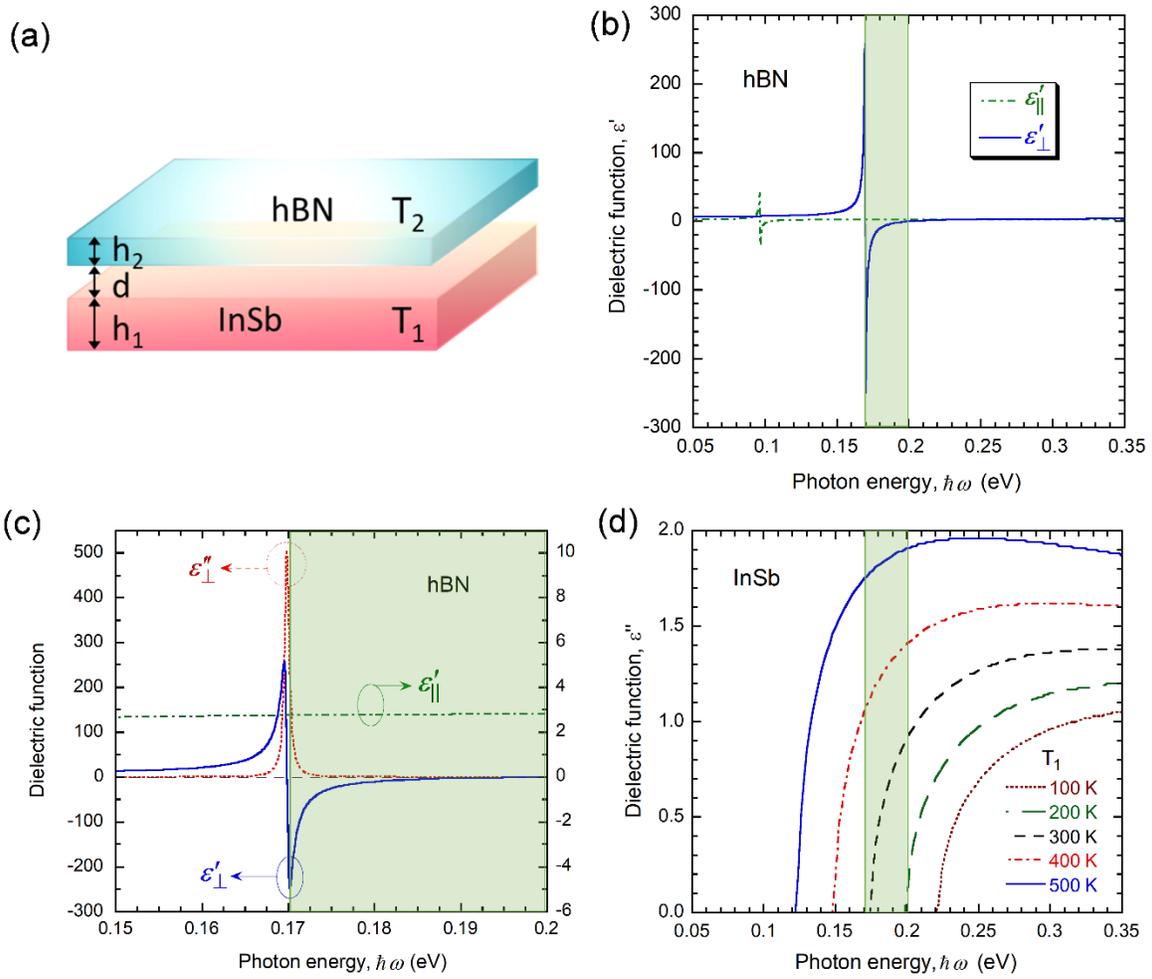

Fig. 1, Feng et al.

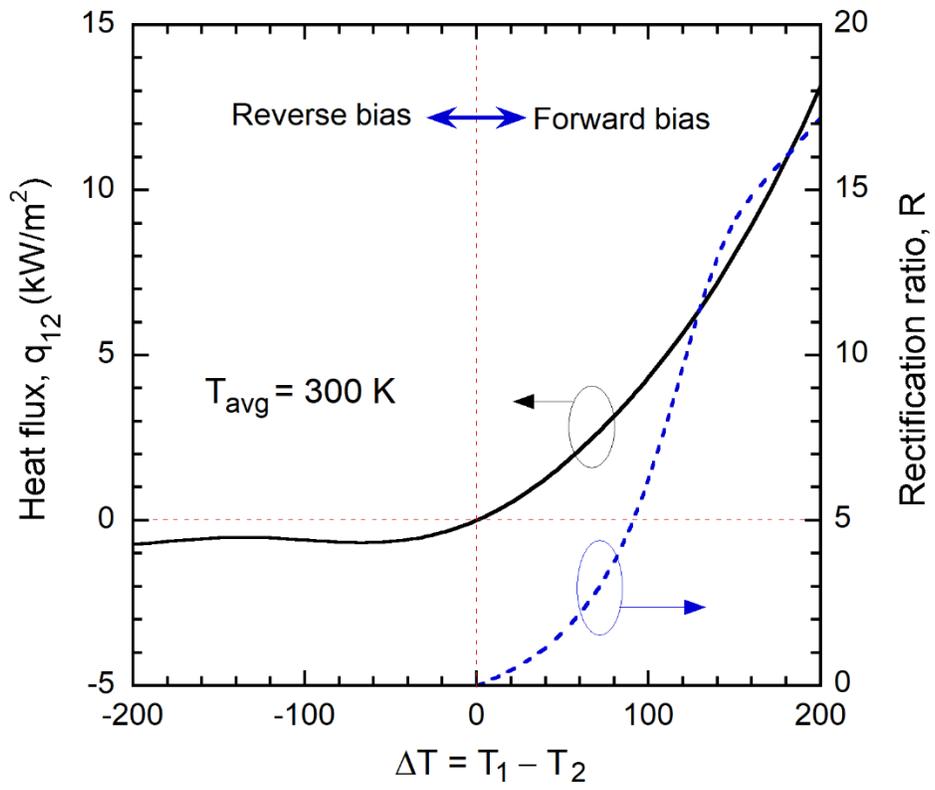

Fig. 2, Feng et al.

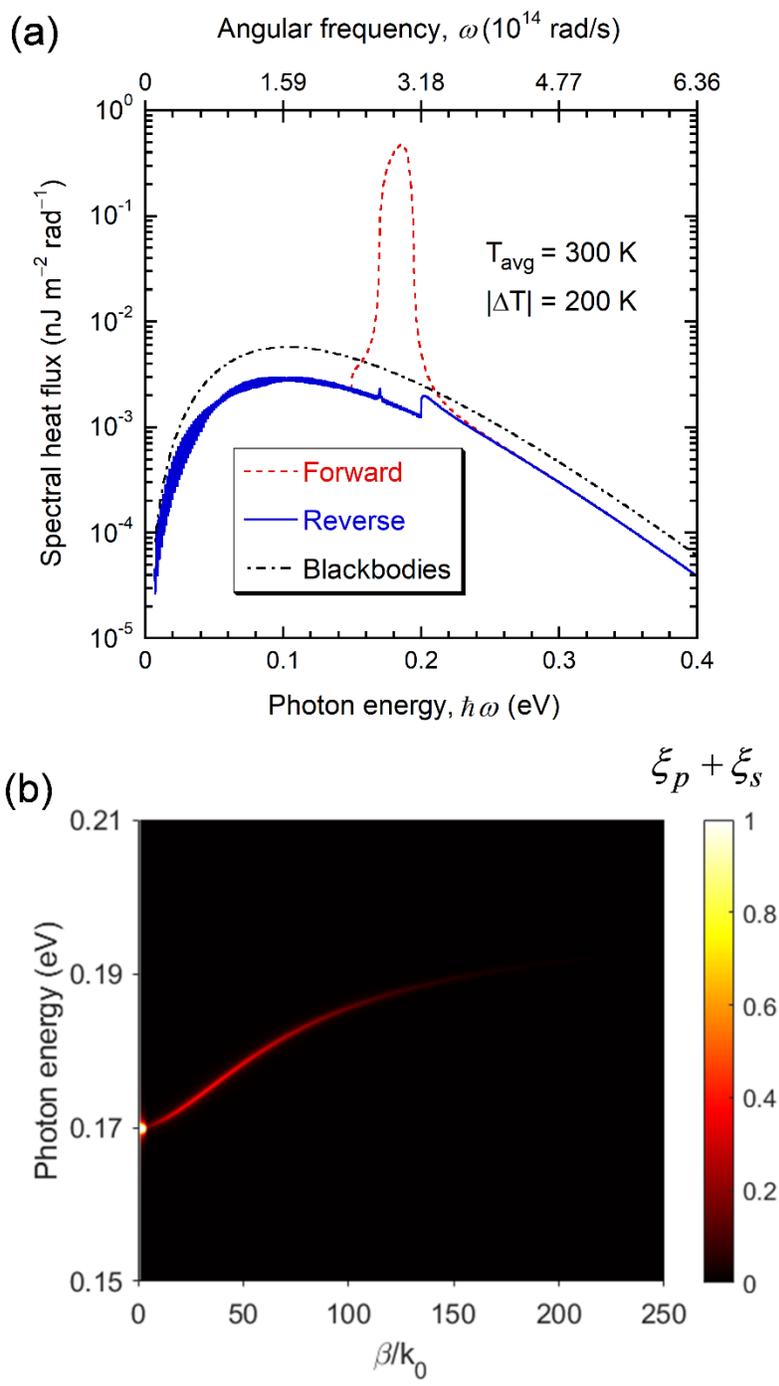

Fig. 3, Feng et al.

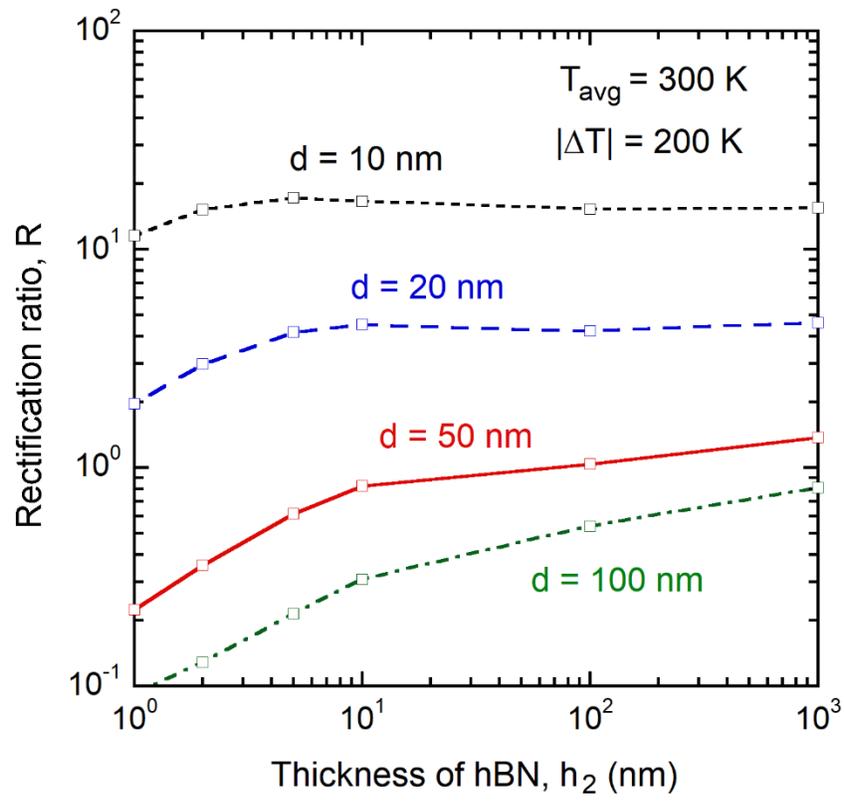

Fig. 4, Feng et al.

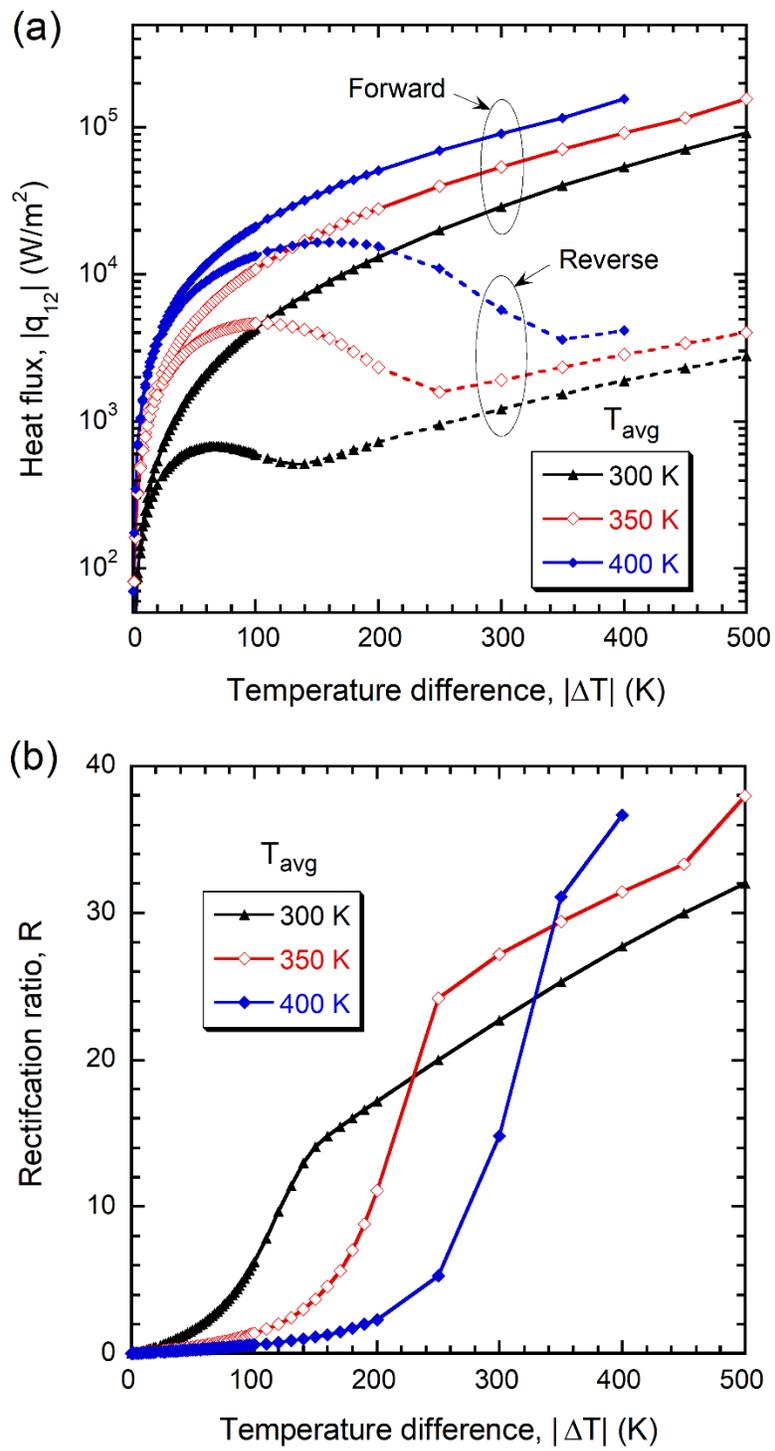

Fig. 5, Feng et al.